\documentclass[titlepage,12pt]{article}
\usepackage{amssymb}
\usepackage{amsfonts}
\begin{document}

{\Large \bf String Theory : a mere prelude to \\ \\
non-Archimedean Space-Time \\ \\
Structures ?} \\ \\

{\bf Elem\'{e}r E Rosinger} \\
Department of Mathematics \\
and Applied Mathematics \\
University of Pretoria \\
Pretoria \\
0002 South Africa \\
eerosinger@hotmail.com \\ \\

{\bf Abstract} \\

It took two millennia after Euclid and until in the early 1880s, when we went beyond the
ancient axiom of parallels, and inaugurated geometries of curved spaces. In less than one more
century, General Relativity followed. At present, physical thinking is still beheld by the yet
deeper and equally ancient Archimedean assumption which entraps us into the limited view of
"only one walkable world". In view of that, it is argued with some rather easily accessible
mathematical support that Theoretical Physics may at last venture into the Non-Archimedean
realms. \\ \\

{\bf 1. A Deep Expression of a Need for Reconsideration} \\

String Theory has for the last nearly three decades known a special status within Theoretical
Physics in the pursuit of the grand unification between General Relativity and Quantum Theory.
Recently however, it has become the object of a considerable criticism due to a number of
reasons which in times to come may, or for that matter, may not turn out to be fully valid. \\

Be it as it may, even in case String Theory ends up side lined for some time to come, or
possibly for good, it is important not to throw away with it those of the {\it more
fundamental novel ideas} which it managed to bring to the forefront of thinking in Theoretical
Physics, ideas that may actually express a deeper and more generally valid need for
reconsideration in the realms of our most fundamental physical intuitions, meanings and
concepts. \\

Two such possible novel ideas worth retaining are the following :

\begin{itemize}

\item seeing space-time as involving a number of {\it dimensions} quite beyond the usual four
space-time ones,

\item seeing the simplest primary elements of space-time {\it not} as mere dimensionless
points, but rather, as entities with a certain structure.

\end{itemize}

Remarkably, the second idea above had been familiar in Mathematics, namely, in Topology, prior
to String Theory. It is what rather whimsically is called {\it pointless topology}, or more
appropriately, {\it point-free topology}, see the early survey [1], with references going back
into the 1960s. \\

As for the first above idea, that as well had been suggested in Physics, prior to String
Theory. Indeed, as far back as in the early 1920s, Kaluza-Klein put forward a five dimensional
version of General Relativity. Following that, still higher dimensional versions of General
Relativity were proposed. \\

In this paper, related to the above two ideas - and at the same time, going beyond them in
ways that have so far mostly escaped the intuition or interest of physicists - we shall draw
attention upon, and argue the possible appropriateness of considering {\it non-Archimedean}
structures for space-time, see Appendix 1 for the respective definition. \\

Here it should be noted that in Theoretical Physics there have been occasional suggestions for
a {\it reconsideration} of our usual perceptions and conceptions of space-time, and
specifically, of its Mathematical modelling, be it geometric, algebraic, topological, and so
on. \\
Interestingly enough, String Theory itself gave importance to what has lately been called
Non-Commutative Geometry. \\
And on yet more basic and simple levels, it has been suggested in Theoretical Physics that the
scalars given by the usual field $\mathbb{R}$ of real numbers, scalars upon which the field
$\mathbb{C}$ of complex numbers itself is built, should be replaced by other sets of scalars,
some of such scalars having a multiplication which is non-commutative, or even non-associative,
see [6-9] and the references cited there. \\
In fact, connected with Quantum Field Theory, there have been suggestions coming from
theoretical physicists, [4], for using N-Category Theory, which is a most involved extension
of Category Theory introduced in Mathematics in the 1940s, the latter itself being already on
such a level of abstraction as to be beyond the present interest of most so called working
mathematicians. \\

A most remarkable latest program for a deep reconsideration of the foundations of Theoretical
Physics has just been published in [1]. Here we cite some of its introductory arguments which
could hardly be expressed better :

\begin{quote}

"A striking feature of the various current programmes for quantising gravity— including
superstring theory and loop quantum gravity—is that, notwithstanding their disparate views on
the nature of space and time, they almost all use more-or-less standard quantum theory.
Although understandable from a pragmatic viewpoint (since all we have is more-or-less standard
quantum theory) this situation is nevertheless questionable when viewed from a wider
perspective. Indeed, there has always been a school of thought asserting that quantum theory
itself needs to be radically changed/developed before it can be used in a fully coherent
quantum theory of gravity.
This iconoclastic stance has several roots, of which, for us, the most important is the use in
the standard quantum formalism of certain critical mathematical ingredients that are taken for
granted and yet which, we claim, implicitly assume certain properties of space and time. Such
an a priori imposition of spatio-temporal concepts would be a major error if they turn out to
be fundamentally incompatible with what is needed for a theory of quantum gravity.
A prime example is the use of the continuum which, in this context, means the real and/or
complex numbers. These are a central ingredient in all the various mathematical frameworks in
which quantum theory is commonly discussed. For example, this is clearly so with the use of
(i) Hilbert spaces and operators; (ii) geometric quantisation; (iii) probability functions on
a non-distributive quantum logic; (iv) deformation quantisation; and (v) formal (i.e.,
mathematically ill-defined) path integrals and the like. The a priori imposition of such
continuum concepts could be radically incompatible with a quantum gravity formalism in which,
say, space-time is fundamentally discrete: as, for example, in the causal set programme."

\end{quote}

What however appears to have avoided the intuition, awareness or interest of physicists is the
use of {\it non-Archimedean} space-time structures. And as argued in [9], the fact that,
mostly by an omission or default, the mathematical models used in Theoretical Physics have
been confined to those which happen to be given by Archimedean structures, has led to an {\it
entrapment} into a so called "one single walkable world", see section 2.4., and entrapment
which no one seems to be aware of in Theoretical Physics. \\
In particular, it is tacitly accepted that time, for instance, extends at most only as far as
- but by {\it no means} beyond - the realms described by the interval of usual real numbers
from $-\infty$ to $+\infty$, that is, as given by the real line $\mathbb{R}$. And such a view
of the ranges of all possible time is taken for granted in Classical Mechanics, Relativity and
Quantum Theory as well. \\
By the same token, it is accepted that in Classical Mechanics and Special Relativity there is
simply {\it no} - and there can {\it never} be any - space whatsoever beyond that which is
described by $\mathbb{R}^3$. \\

To a certain extent, such tacit and rather universal assumptions have an historical
explanation - however, not necessarily and excuse as well - in the fact that the field
$\mathbb{R}$ of real numbers is the only field which is linearly ordered, complete, and
Archimedean. And at the beginnings of ancient Mathematics, and specifically Geometry, there
was obviously no use in any non-Archimedean structures. \\

Certainly, at its ancient origins, Mathematics was confined to the realms directly accessible
to our human senses, that is the realms of the physical, and specifically the visible and the
palpable. And quantitative measurement of various objects, and also of land, was among its
main concerns and applications. In such a context, no doubt, the Archimedean assumption had to
arise. After all, without it, there would not be any possibility to choose a unit measure, and
then measure by its integer and fractional multiples the visible and palpable entities under
consideration. \\

Remarkably, ever since ancient times and till recently, the Archimedean assumption has never
been questioned even in Mathematics, let alone in Physics. On the contrary, it went on and on
as an assumption on a level still deeper than those formulated explicitly as various
self-evident axioms. \\
It is therefore not so surprising that Theoretical Physics still treats the Archimedean
assumption precisely in the same manner. \\

In this regard, it is rather poignant to recall the development of the views related to the
ancient axiom of parallels. \\
Until the early 1800s, no one ever questioned the veracity of that axiom. On the contrary, it
was assumed to be true, and all efforts were directed towards proving it from the other axioms.
Indeed, the only concern about the axiom of parallels was caused by the fact that, when
compared with the other axioms of Geometry, it seemed somewhat complex, and also, it involved
the concept of infinity, a concept not directly accessible to our human senses. Thus it was
the common perception that the axiom of parallels could not so easily be classified along the
other axioms as self-evident. \\
And until the early 1800s, that was an obvious matter of concern, since it was taken for
granted that :

\begin{quote}

"Axioms can only and only be self-evident statements.

\end{quote}

Here in this last statement itself we can see once again a clear and relevant example of the
crucial importance of certain {\it axioms which are accepted on a deeper, and less than
conscious level}. Let us call them for convenience {\it deep axioms}. And as it often happens,
such deep axioms can easily play a more important role than the usual ones which we are
consciously and explicitly formulated and accepted. \\

And how deep the axiom of parallels managed to penetrate in human awareness is illustrated by
the fact that not long before the discovery of non-Euclidean Geometries in which that axiom
does no longer hold, such a remarkable philosopher like Immanuel Kant, distinguished with an
incisive critical sense, kept considering Euclidean Geometry, and thus the axiom of parallels,
as nothing short of being an a priori truth, therefore being the {\it only} possible
Geometry. \\

Fortunately, in the early 1800s, the axiom of parallels was proved to be independent of the
other axioms of Geometry. Furthermore, two different versions of that axiom could clearly be
formulated, thus leading to two non-Euclidean Geometries. \\
Needless to say, without that discovery, General Relativity, which got discovered in less than
one more century, could not have come about.

And so it came to pass that within a few generations, the questioning of the axiom of
parallels opened the way to such a revolution in Theoretical Physics, as that brought about by
General Relativity. \\

Yet, today, the Archimedean assumption still keeps physicists enthralled ... \\
Enthralled by, and also entrapped into "one single walkable world" which it inevitably
imposes ... \\
And the resulting limitations, of which no one seems to be aware of in Theoretical Physics,
can be tremendous. For instance, in a non-Archimedean space-time structure the very concept of
quantum, as well as the finite upper limit on possible velocities cannot but acquire
completely new meanings, and possibly, even new formulations or alternatives ... \\

However, several modern facts of a rather historic proportion in themselves, should have by
now drawn attention upon our entrapment into "one single walkable world" with its Archimedean
limitations. Among these facts is the following one : \\

Back in 1966, Abraham Robinson introduced Nonstandard Analysis. His reason was mainly to place
on a rigorous footing the "infinitesimals" used by Leibniz in Calculus, in the late 1600s.
However, the point of importance here in Robinson's construction of the field $^*\mathbb{R}$
of nonstandard reals is the {\it non-Archimedean} nature of that field, a property which must
of course follow from the uniqueness of the Archimedean field $\mathbb{R}$ of real numbers
mentioned above. \\

To this crucial fact, and being by no means less important, one can add the most intriguing
question raised in [5], which can briefly be formulated as follows :

\begin{quote}

How come that all spaces used so far in Theoretical Physics have as sets a cardinal {\it not}
larger than that of the continuum, that is, of the real numbers $\mathbb{R}$ ?

\end{quote}

And the relevance of this question is in the fact that, ever since Cantor's Set Theory,
introduced in the late 1800s, we know about sets with cardinality incomparably larger that of
the continuum. Furthermore, the cardinal of the continuum is very low among the infinite
cardinals. In fact, it is merely the second one, namely, after that of the cardinal of the set
$\mathbb{N}$ of nonnegative integers, if one accepts the Continuum Hypothesis. \\

However, what may be no less important an argument than those above for the need of a {\it
reconsideration} of the Archimedean mathematical structures used in Theoretical Physics are
the surprising and so far not yet explored rich opportunities opened up by non-Archimedean
structures. \\

A well developed example about the actual benefits of such considerable opportunities is given,
among others, by the Nonlinear Algebraic Theory of Generalized Solution for large classes of
nonlinear partial differential equations, a theory originated in the early 1960s, and by now
listed by the American Mathematical Society in their Subject Classification, under 46F30,
see : \\

www.ams.org/msc/46Fxx.html \\

Some of the relevant such applications of non-Archimedean structures, and not only in the
solution of partial differential equations, can be found in [10-33], and in the references
cited there. \\

Returning to [5], my general appreciation of it, expressed succinctly and without much detail,
can be found in Appendix 2 which was sent for publication to The Mathematical Intelligencer. \\
Regardless of that view, however, one can note that [5] does not give any attention to the
issue of Archimedean versus non-Archimedean structures, and instead, and merely by default,
takes the traditional Archimedean view for granted. \\

Recently, related to Quantum Gravity, so called {\it background free} theoretical models have
been suggested. \\
As it happens, however, such models are still entrapped into the "one single walkable world"
situation, even if they see themselves free from any usual Geometry, Algebra or Topology. And
this entrapment can only one again show how deep some deep axioms can indeed condition one's
whole vision and thinking, even if such deep axioms are by then present only by their, so to
say, ghosts ... \\

The ancient axiom of parallels created a long ongoing controversy which got only solved in the
early 1880s. \\
And this solution, in less than one more century, opened the way to General Relativity. \\

On the other hand, the Archimedean assumption has so far hardly brought with it any comparable
controversy. And certainly not in Theoretical Physics ... \\
And as things so often go with us humans : no controversy means no consideration, let alone
reconsideration ... \\
Thus in Theoretical Physics we remain trapped for evermore into the Archimedean
assumption and the corresponding "one single walkable world" which it imposes ... \\

Going beyond the ancient axiom of parallels was, no doubt, absolutely {\it necessary} for
being able to conceive of and achieve General Relativity. \\
Fortunately, once the transition beyond that ancient axiom happened, in less than one more
century, it also proved to be {\it sufficient} for that truly revolutionary success in
Theoretical Physics to occur. \\

Nowadays, it is perhaps the time to go beyond the yet deeper ancient Archimedean assumption,
and at long last, liberate physical thinking from the limitations of the consequent "one
single walkable world" ... \\ \\

{\bf 2. Universes within Universes, Universes next to Universes ...} \\

\medskip
We shall briefly illustrate here the surprising wealth of structure the non-Archimedean
property can bring with itself. A few further details in this regard can be found in [9]. \\

\medskip
{\bf 2.1. A simple basic one dimensional and linearly ordered \\
          \hspace*{0.8cm} example} \\

For simplicity, we start with one of the most relevant non-Archimedean fields which extend the
usual scalar field $\mathbb{R}$ of real numbers, namely, the field $^*\mathbb{R}$ of
nonstandard real numbers, introduced by Robinson in the 1960s. And to make more {\it user
friendly} the presentation of the facts important here - namely, the surprising richness of
the respective non-Archimedean structure - we shall recall the corresponding results without
proof, since as is well known, such proofs, presented for instance in [3], can be technically
rather involved, as it is typical for Nonstandard Analysis. \\

Let us therefore see

\begin{itemize}

\item what are the differences between the Archimedean field of usual real scalars in
$\mathbb{R}$, and  on the other hand, their non-Archimedean field extension given by
$^*\mathbb{R}$ ?

\item how much more rich is the structure of $^*\mathbb{R}$, when compared with that of
$\mathbb{R}$ ?

\end{itemize}

As we shall see in the sequel, going from $\mathbb{R}$ to the much larger $^*\mathbb{R}$
involves {\it two expansions}, namely, a {\it local} one, at each point $r \in \mathbb{R}$, as
well as a {\it global} one, at both infinite ends of the real line $\mathbb{R}$. \\

A particularly useful way to illustrate it, [3], is by saying that, when we want to go from
$\mathbb{R}$ to $^*\mathbb{R}$, we need {\it two special instruments} which can give us views
outside of $\mathbb{R}$, and into the not yet seen, and not yet known worlds situated within
$^*\mathbb{R}$ beyond the confines of the standard realm of the real line $\mathbb{R}$. Namely,
we need

\begin{itemize}

\item a so called {\it microscope} in order to see the {\it monads} which give the new and
additional {\it local structure} in $^*\mathbb{R}$, and we also need

\item a so called {\it telescope} for being able to look at {\it galaxies} giving the new and
additional {\it global structure} in $^*\mathbb{R}$.

\end{itemize}

In order to understand the structure of $^*\mathbb{R}$, the following three things are
therefore enough :

\begin{itemize}

\item to keep in mind that $^*\mathbb{R}$ is a {\it linearly} ordered non-Archimedean field,

\item to understand how the {\it monads} create the {\it local} structure of $^*\mathbb{R}$,
and in this respect, it is enough to understand how they create the local structure of the
{\it galaxy} of $0 \in \mathbb{R}$, which can be seen as the central galaxy,

\item to understand how by uncountably many translations to the right and left, the galaxy of
$0 \in \mathbb{R}$, that is, the central galaxy, creates the {\it global} structure of
$^*\mathbb{R}$.

\end{itemize}

Anticipating and simplifying a more precise description which follows in the sequel, the way
$^*\mathbb{R}$ is obtained from $\mathbb{R}$ is in summary by the next procedure :

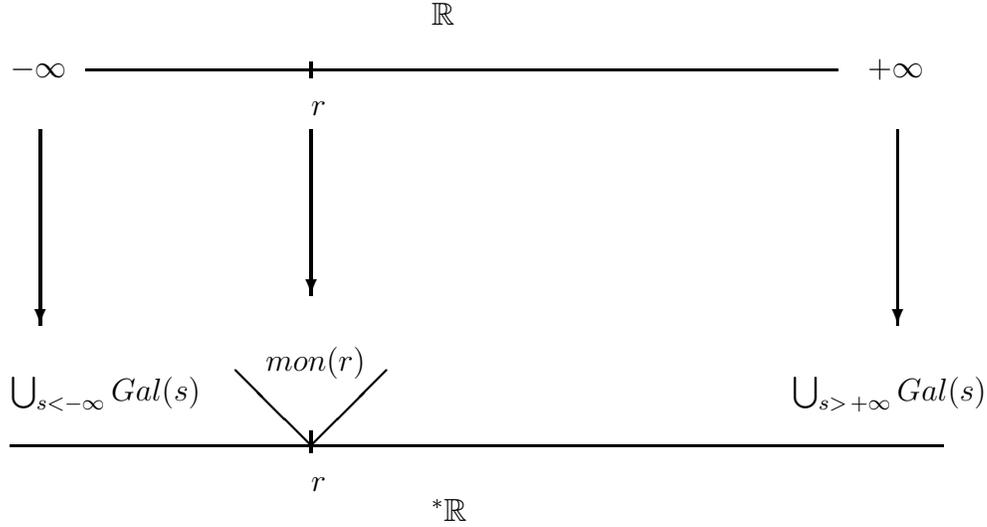
\begin{math}
\setlength{\unitlength}{0.2cm}
\thicklines
\begin{picture}(65,38)

\put(28,33){$\mathbb{R}$}
\put(0,29.5){$- \infty$}
\put(5,30){\line(1,0){50}}
\put(20,29.5){\line(0,1){1}}
\put(57,29.5){$+ \infty$}
\put(20,27){$r$}
\put(2,26){\vector(0,-1){13}}
\put(0,8){$\bigcup_{s < -\infty} Gal ( s )$}
\put(20,26){\vector(0,-1){11}}
\put(17,10){$mon ( r )$}
\put(20,5){\line(-1,1){5}}
\put(20,5){\line(1,1){5}}
\put(20,4.5){\line(0,1){1.5}}
\put(59,26){\vector(0,-1){13}}
\put(52,8){$\bigcup_{s >\, +\infty} Gal ( s )$}
\put(0,5){\line(1,0){62}}
\put(20,2){$r$}
\put(28,0){$^*\mathbb{R}$}

\end{picture}
\end{math} \\

In other words, $^*\mathbb{R}$ is obtained from $\mathbb{R}$ as follows :

\begin{itemize}

\item at each usual real number $r \in \mathbb{R}$ there is an {\it expansion} by the
insertion of the uncountable set $mon( r )$, furthermore

\item at each of the $\pm \infty$ ends of $\mathbb{R}$, there is an {\it expansion} in which
uncountable unions of uncountable sets $Gal ( s )$ are {\it joined} to $\mathbb{R}$.

\end{itemize}

Keisler's microscope is needed in order to be able to look into $mon ( 0 )$, since the
infinitesimals cannot be seen from the usual point of view of $\mathbb{R}$. This also goes of
course for each $mon ( r )$, with $r \in \mathbb{R}$. \\
In this way, in order to see every number in $Gal ( 0 )$, we need Keisler's microscope. \\

Similarly, Keisler's telescope is needed in order to see what is going on to the left of
$-\infty$, and to the right of $+\infty$. Indeed, galaxies other than $Gal ( 0 )$, we cannot
see without Keisler's telescope. \\

And now, let us make the above more precise by giving some information on the structure of the
uncountable sets $mon ( )$ and $Gal ( )$. \\

A scalar $s \in\, ^*\mathbb{R}$ is called {\it infinitesimal}, if and only if $| s | \leq r$,
for every $r \in \mathbb{R},~ r > 0$. The set of such infinitesimal scalars is denoted by \\

$~~~~~~ mon ( 0 ) $ \\

and following Leibniz, it is called the {\it monad} of $0 \in \mathbb{R}$. \\

For an arbitrary scalar $s \in\, ^*\mathbb{R}$ it will be convenient to denote \\

$~~~~~~ mon ( s ) = s + mon ( 0 ) $ \\

which is but the translate of $mon ( 0 )$ by $s$, and it is the {\it monad} of $s$. \\

A scalar $s \in\, ^*\mathbb{R}$ is called {\it finite}, if and only if $| s | \leq r$, for
some $r \in \mathbb{R},~ r > 0$. We denote by \\

$~~~~~~ Gal ( 0 ) $ \\

the set of all finite scalars , and call that set the {\it galaxy} of $0 \in \mathbb{R}$, or
the {\it central} galaxy, since it is the only glaxy which contains the real line
$\mathbb{R}$. \\

For an arbitrary scalar $s \in\, ^*\mathbb{R}$ it will be convenient to denote \\

$~~~~~~ Gal ( s ) = s + Gal ( 0 ) $ \\

which is but the translate of $Gal ( 0 )$ by $s$, and it is the {\it galaxy} of $s$. \\

At last, a scalar $s \in\, ^*\mathbb{R}$ is called {\it infinite}, if and only if $| s | \geq
r$, for every $r \in \mathbb{R},~ r > 0$. \\

As follows easily in Nonstandard Analysis, [3], none of the above sets of monads and galaxies
is void, and in fact, each of them is uncountably large. Furthermore, as seen in the sequel,
both monads and galaxies have surprisingly complex structures which in a way {\it reflect} one
another, and in addition, each of them is {\it self-similar}, recalling one of the well known
basic properties of {\it fractal} structures. \\

Now we recall that  \\

(2.1)~~~ $ mon ( 0 ) ~\bigcap~ \mathbb{R} ~=~ \{~ 0 ~\} $ \\

thus the only real number $r \in \mathbb{R}$ which is infinitesimal is $r = 0 \in
\mathbb{R}$. Also, for $x,~ y \in \mathbb{R}$ and $x \neq y$, we have \\

(2.2)~~~ $ mon ( x ) ~\bigcap~ mon ( y ) ~=~ \phi $ \\

hence we obtain the representation given by a union of {\it pairwise disjoint} sets \\

(2.3)~~~ $ Gal ( 0 ) ~=~ \bigcup_{r \,\in\, \mathbb{R}} mon ( r ) $ \\

Further we have \\

(2.4)~~~ $ \mathbb{R} ~\subsetneqq~ Gal ( 0 ) ~\subsetneqq~ ^*\mathbb{R} $ \\

(2.5)~~~ $ ^*\mathbb{R} ~\setminus~ Gal ( 0 ) $~~ is the set of infinite numbers in
$^*\mathbb{R}$ \\

It follows that, schematically, we have for $^*\mathbb{R}$

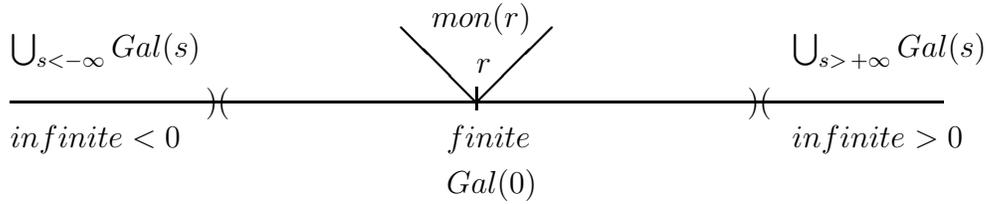
\begin{math}
\setlength{\unitlength}{0.2cm}
\thicklines
\begin{picture}(65,15)

\put(0,8){$\bigcup_{s < -\infty} Gal ( s )$}
\put(28,10){$mon ( r )$}
\put(31,5){\line(-1,1){5}}
\put(31,5){\line(1,1){5}}
\put(31,4.5){\line(0,1){1.5}}
\put(52,8){$\bigcup_{s >\, +\infty} Gal ( s )$}
\put(0,5){\line(1,0){62}}
\put(13,4.4){$)($}
\put(49,4.4){$)($}
\put(31,7){$r$}
\put(0,2){$infinite < 0$}
\put(29,2){$finite$}
\put(52,2){$infinite > 0$}
\put(29,-1){$Gal ( 0 )$}

\end{picture}
\end{math} \\

And similar with (2.2), for $s, t \in \,^*\mathbb{R}$ and $t - s$ infinite, we have \\

(2.6)~~~ $ Gal ( s ) ~\bigcap~ Gal ( t ) ~=~ \phi $ \\

while for $t - s$ finite, we have \\

(2.7)~~~ $ Gal ( s ) ~=~ Gal ( t ) $ \\

As a consequence, and similar with (2.3), we have the following representation as a union of
{\it pairwise disjoint} sets \\

(2.8)~~~ $ ^*\mathbb{R} ~=~ \bigcup_{\lambda \,\in\, \Lambda} Gal ( -s_\lambda )
               ~\bigcup~ Gal ( 0 ) ~\bigcup_{\lambda \,\in\, \Lambda} Gal ( s_\lambda ) $ \\

where $\Lambda$ is an uncountable set of indices, while $s_\lambda \in \,^*\mathbb{R}$ are
positive and infinite for $\lambda \in \Lambda$, and such that $s_\lambda - s_\mu$ is infinite
for $\lambda, \mu \in \Lambda,~ \lambda \neq \mu$. \\

The use of the term {\it monad} by Robinson was, as mentioned, inspired by Leibniz who first
employed infinitesimals and did so in surprisingly effective manner, even if in a merely
intuitive and insufficiently rigorous manner when, parallel with and independently of Newton,
started the development of Calculus in the late 1600s. In this regard the fact that \\

$~~~~~~ mon ( 0 ) ~\bigcap~ \mathbb{R} ~=~ \{~ 0 ~\} $ \\

which means that zero is the {\it only} real number which is infinitesimal, caused, starting
with Leibniz, all sort of difficulties when, prior to the creation of modern Nonstandard
Analysis, one tried to deal with infinitesimals. Similarly, the fact that \\

$~~~~~~ ^*\mathbb{R} ~\setminus~ Gal ( 0 ) $ \\

is the set of infinite numbers in $^*\mathbb{R}$, thus there are {\it no} reals numbers which
are infinite, brought with it difficulties when dealing with infinitely large numbers, prior
to the modern theory of Nonstandard Analysis. \\

\medskip
{\bf 2.2. Global and local structure of $^*\mathbb{R}$} \\

The structure of $Gal ( 0 )$, both {\it locally} and {\it globally}, is presented in (2.3). In
view of (2.8), that also gives an understanding of the {\it local} structure of
$^*\mathbb{R}$. \\

For the understanding of the {\it global} structure of $^*\mathbb{R}$ we have the {\it order
reversing bijective} mapping between infinite and infinitesimal numbers in $^*\mathbb{R}$,
namely \\

(2.9)~~~ $ (~ ^*\mathbb{R} ~\setminus~Gal ( 0 ) ~) \ni s ~\longmapsto~
                         1 / s \in (~ mon ( 0 ) \setminus \{~ 0 ~\} ~) $ \\

Indeed, it is obvious that, given $s, t \in\, ^*\mathbb{R} ~\setminus~Gal ( 0 )$, then \\

$~~~~~~ s ~<~ t ~~~\Longleftrightarrow~~~ 1 / s ~>~ 1 / t $ \\

This fact is fundamental, as it shows that the local structure of $^*\mathbb{R}$ {\it mirrors}
its global structure, and vice versa. \\

It also establishes the link between Keisler's microscope and telescope, the former letting us
see into $mon ( 0 )$, while the latter allowing us to look out into $^*\mathbb{R} ~\setminus~
Gal ( 0 )$. \\

In this regard, the properties (2.1) - (2.7) express the {\it local} structure of
$^*\mathbb{R}$, more precisely, within its part given by $Gal ( 0 )$, that is, the structure
of the finite numbers in $^*\mathbb{R}$. This local structure consists of a sort of
infinitesimal neighbourhood around every usual real number $r \in \mathbb{R}$, neighbourhood
given by the translate ~$mon ( r ) = r ~+~ mon ( 0 )$~ of ~$mon ( 0 )$. And in view of (2.1),
this local structure cannot be seen from $\mathbb{R}$, thus we have to use Keisler's
microscope. \\

\medskip
{\bf 2.3 Self-Similarity in $^*\mathbb{R}$} \\

The local and global structures of $^*\mathbb{R}$, as we have seen in subsection 2.2, are
closely related, as they are expressed in (2.3), (2.8) and (2.9). \\

Here we shall point to a {\it self-similar} aspect of this interrelation which may remind us
of a typical feature of fractals. \\

In this regard, we recall that the global structure of $^*\mathbb{R}$ is given by, see (2.8) \\

$~~~~~~ ^*\mathbb{R} ~=~ \bigcup_{\lambda \,\in\, \Lambda} Gal ( -s_\lambda )
               ~\bigcup~ Gal ( 0 ) ~\bigcup_{\lambda \,\in\, \Lambda} Gal ( s_\lambda ) $ \\

while its local structure is described by, see (2.9) \\

$~~~~~~ Gal ( 0 ) ~=~ \bigcup_{~r \in \mathbb{R}}~ (~ r ~+~ mon ( 0 ) ~)$ \\

In this way we obtain the {\it self-similar order reversing bijections}, which are expressed
in terms of $mon ( 0 )$, namely

\begin{math}
\setlength{\unitlength}{0.2cm}
\thicklines
\begin{picture}(65,19)

\put(0,15){$(2.10)$}
\put(0,10){$[\, \bigcup_{r \in \mathbb{R},\, \lambda \in \Lambda}\,
                          ( r - s_\lambda + mon ( 0 ) ) \,] \cup
                   [\, \bigcup_{r \in \mathbb{R},\, \lambda \in \Lambda}
                          ( r + s_\lambda + mon ( 0 ) ) \,] \ni s ~\longmapsto$}
\put(2,3){$\longmapsto~ 1 / s \in (~ mon ( 0 ) \setminus \{~ 0 ~\} ~)$}

\end{picture}
\end{math}

and conversely

\begin{math}
\setlength{\unitlength}{0.2cm}
\thicklines
\begin{picture}(65,19)

\put(0,15){$(2.11)$}
\put(0,10){$(~ mon ( 0 ) \setminus \{~ 0 ~\} ~) \ni s \longmapsto$}
\put(2,3){$1 / s \in [\, \bigcup_{r \in \mathbb{R},\, \lambda \in \Lambda}\,
                          ( r - s_\lambda + mon ( 0 ) ) \,] \cup
                   [\, \bigcup_{r \in \mathbb{R},\, \lambda \in \Lambda}
                          ( r + s_\lambda + mon ( 0 ) ) \,]$}
\end{picture}
\end{math}

As we can note, the above bijections in (2.10), (2.11) are given by the very simple algebraic,
explicit, and order reversing mapping $s \longmapsto 1 / s$ which involves what is essentially
a field operation, namely, division. And these two bijections take the place of the much
simpler order reversing bijections in the case of the usual real line $\mathbb{R}$, namely \\

(2.12)~~~ $ (~ \mathbb{R} \setminus ( -1 , 1 ) ~) \ni r ~\longmapsto~
                          1 / r \in (~ [ - 1, 1 ] \setminus \{~ 0 ~\} ~) $ \\

(2.13)~~~ $ (~ [ - 1, 1 ] \setminus \{~ 0 ~\} ~) \ni r ~\longmapsto~
                          1 / r \in (~ \mathbb{R} \setminus ( -1 , 1 ) ~) $ \\

The considerable difference between (2.10), (2.11), and on the other hand, (2.12), (2.13) is
obvious. In the former two, which describe the structure of $^*\mathbb{R}$, the order
reversing bijections represent the set \\

$~~~~~~ mon ( 0 ) \setminus \{~ 0 ~\} $ \\

through the set \\

$~~~~~~ [\, \bigcup_{r \in \mathbb{R},\, \lambda \in \Lambda}\,
                          ( r - s_\lambda + mon ( 0 ) ) \,] \cup
                   [\, \bigcup_{r \in \mathbb{R},\, \lambda \in \Lambda}
                          ( r + s_\lambda + mon ( 0 ) ) \,] $ \\

which contains uncountably many translates of the set $mon ( 0 )$. And it is precisely this
manifestly involved {\it self-similarity} of the set $mon ( 0 )$ of {\it monads} which is the
{\it novelty} in the non-Archimedean structure of $^*\mathbb{R}$, when compared with the much
simpler Archimedean structure of $\mathbb{R}$. This novelty is remarkable since it makes $mon
( 0 )$ having the same complexity with \\

$~~~~~~ ^*\mathbb{R} \setminus Gal ( 0 ) ~=~ $ \\

$~~~~~~~~~~~~=~ [\, \bigcup_{r \in \mathbb{R},\, \lambda \in \Lambda}\,
                          ( r - s_\lambda + mon ( 0 ) ) \,] \cup
                   [\, \bigcup_{r \in \mathbb{R},\, \lambda \in \Lambda}
                          ( r + s_\lambda + mon ( 0 ) ) \,] $ \\

In this way $mon ( 0 )$, which is but the realm of {\it infinitesimals}, thus it {\it cannot}
be seen in terms of $\mathbb{R}$, except with Keisler's microscope, has the very same
complexity as the set $^*\mathbb{R} \setminus Gal ( 0 )$ of all infinitely large numbers,
which again {\it cannot} be seen from $\mathbb{R}$, unless Keisler's telescope is used. \\

As for (2.12), (2.13), the utter simplicity of the Archimedean trap becomes obvious, since the
sets $[ - 1, 1 ] \setminus \{~ 0 ~\}$ and $\mathbb{R} \setminus ( - 1, 1 )$ are again in an
order reversing bijection. However, no any trace whatsoever of self-similarity, since $[ - 1,
1 ]$ does not even once appear in $\mathbb{R} \setminus ( - 1, 1 )$. \\

\medskip
{\bf 2.4. Walkable worlds ...} \\

The local and global nature of a non-Archimedean structure can be illustrated in a more
intuitively accessible and geometric manner, the manner which in ancient times was quite
likely the one taken in view of its obvious everyday practicality. And to keep things simple,
we remain for a while longer with the nonstandard extension $^*\mathbb{R}$ of the real line
$\mathbb{R}$. Indeed, $^*\mathbb{R}$ is a highly relevant instance of a non-Archimedean
structure, thus it is in this regard essentially different from $\mathbb{R}$ which is
Archimedean. On the other hand, both $^*\mathbb{R}$ and $\mathbb{R}$ have the simplicity of
being one dimensional. \\

Given now any usual real number $u \in \mathbb{R},~ u > 0$, let as ask what appears to be an
eminently practical question, namely :

\begin{quote}

How far along the one dimensional world of $^*\mathbb{R}$ can one walk from any given point $r
\in \mathbb{R}$ by taking an arbitrary large finite number of steps of size $u$, be it in the
positive or negative directions ?

\end{quote}

The answer given by Nonstandard Analysis is \\

(2.14)~~~ $ Gal ( 0 ) $ \\

Needless to say, in view of (2.8), this kind of walk confines us to an utterly small part of
$^*\mathbb{R}$. Namely, as follows from (2.3), $Gal ( 0 )$ only contains the usual real line
$\mathbb{R}$ and the monads around each real number $r \in \mathbb{R}$, thus the infinitesimal
neighbourhoods of such numbers $r$. Consequently, $Gal ( 0 )$ does not contain any of the
infinite numbers in $^*\mathbb{R}$. \\

And as is so far mostly the case, it is precisely this limited world which is the one
dimensional world of Theoretical Physics. And in fact it is even less, since it is only its
strict subset given by the real line $\mathbb{R}$, and without any of the monads which make up
the difference between $Gal ( 0 )$ and the smaller $\mathbb{R}$, see (2.3). Indeed, this is
the

\begin{quote}

"only one walkable world"

\end{quote}

within which the mathematical modelling of Physics tends to be confined ... \\

And then, let us see

\begin{quote}

How many other, and even more importantly, what kind of other "walkable worlds" can the
non-Archimedean structure of $^*\mathbb{R}$ offer us ?

\end{quote}

And here may quite likely be a novel and richly promising opportunity to be made use of at
last in Theoretical Physics. \\

The answer to the above general question is very simple. We take an arbitrary step length $u
\in\, ^*\mathbb{R},~ u > 0$ and an arbitrary starting point $s \in\, ^*\mathbb{R}$, and denote
by \\

(2.15)~~~ $ WW_{u,\,s} $ \\

that part of $^*\mathbb{R}$ which can be reached from $s$ by walking any finite number of
steps, each of the given length $u$, and do so either in the positive, or in the negative
direction. We call $WW_{u,\,s}$ the "walkable world" from $s$ and with steps $u$. \\

To be more precise mathematically, any given $t \in\, ^*\mathbb{R}$ belongs to $WW_{u,\,s}$\,,
if and only if there exists $n \in \mathbb{N},~ n \geq 1$, such that \\

(2.16)~~~ either ~~$ s ~\leq~ t ~\leq~ s + n u $~~ or ~~$ s - n u ~\leq~ t ~\leq~ s $ \\

or equivalently \\

(2.16$^*$)~~~ $ s - n u ~\leq~ t ~\leq~ s + n u $ \\

An immediate consequence is that \\

(2.17)~~~ all "walkable worlds" are order isomorphic with ~~$Gal ( 0 ) $ \\

Indeed, with the above notation, let us define the mapping $\omega$ by \\

$~~~~~~ WW_{u,\,s} \ni t ~\longmapsto~ \omega ( t ) ~=~ ( t - s ) / u \in\, ^*\mathbb{R} $ \\

then $\omega$ is obviously injective and strictly increasing, while (2.16$^*$) implies that
$\omega ( WW_{u,\,s} ) \subseteq Gal ( 0 )$. However, $\omega$ is also surjective onto $Gal
( 0 )$, since for given $v \in Gal ( 0 )$, we have $-n \leq v \leq n$, for a suitable $n \in
\mathbb{N},~ n \geq 1$, thus $t = s + v u \in WW_{u,\,s}$\,,, while obviously $\omega ( t ) =
v$. \\

Clearly therefore : \\

(2.18)~~~ $ \begin{array}{l}
                \mbox{A "walkable world" {\it cannot} have a smallest or} \\
                \mbox{a largest element.}
             \end{array} $ \\

Another immediate consequence is the following {\it self-confinement} property of such
"walkable worlds", namely \\

(2.19)~~~ $ \forall~~ t \in WW_{u,\,s} ~:~ WW_{u,\,t} ~=~ WW_{u,\,s} $ \\

In other words, one can never leave, or get out of a "walkable world" $WW_{u,\,s}$\,, no
matter at what point $t$ in it one would start walking, and doing so with steps of the same
length $u$, or for that matter, with steps of any length $u\,' = c u$, where $c > 0$ is
finite. \\

In particular it follows that, see (2.14) \\

(2.20)~~~ $ \forall~~ u \in \mathbb{R},~ u \geq 1,~ r \in \mathbb{R} ~:~
                                                       WW_{u,\,r} ~=~ Gal ( 0 ) $ \\

which, in the case of one dimension, contains what is so far the typical "only one walkable
world" in Theoretical Physics ... \\

\medskip
{\bf 2.5. How are the walkable worlds situated with respect to \\
          \hspace*{0.8cm} one another ?} \\

The answer to this question is obtained easily. Suppose given two "walkable worlds"
$WW_{u,\,s}$ and $WW_{u\,',\,s\,'}$, with $u,~ u\,',~ s,~ s\,' \in\, ^*\mathbb{R},~ u,~ \\
u\,' > 0$, and let us assume that these two "walkable worlds" are not disjoint, namely \\

(2.21)~~~ $ WW_{u,\,s} \bigcap WW_{u\,',\,s\,'} ~\neq~ \phi $ \\

Thus we can take a point \\

$~~~~~~ t \in WW_{u,\,s} \bigcap WW_{u\,',\,s\,'} $ \\

and then in view of (2.16$^*$), for suitable $n,~ n\,' \in \mathbb{N},~ n,~ n\,' \geq 1$, we
obtain the inequalities \\

$~~~~~~ s - n u ~\leq~ t ~\leq~ s + n u,~~~
                  s\,' - n\,' u\,' ~\leq~ t ~\leq~ s\,' + n\,' u\,' $ \\

which imply that \\

(2.22)~~~ $ s - n u ~\leq~ s\,' + n\,' u\,',~~~ s\,' - n\,' u\,' ~\leq~ s + n u $ \\

Now since $^*\mathbb{R}$ is linearly ordered, we can assume without loss of generality that \\

(2.23)~~~ $ 0 ~<~ u ~\leq~ u\,' $ \\

and then the first inequality in (2.22) gives \\

$~~~~~~ s ~\leq~ s\,' + n u + n\,' u\,' ~\leq~ s\,' + ( n + n\,' ) u\,' $ \\

hence $s \in WW_{u\,',\,s\,'}$, which in view of (2.23) means that \\

(2.24)~~~ $ WW_{u,\,s} ~\subseteq~ WW_{u\,',\,s\,'} $ \\

In this way, we proved that : \\

(2.25)~~~ $ \begin{array}{l}
                    \mbox{Two "walkable worlds" are either disjoint,} \\
                    \mbox{or one contains the other.}
             \end{array} $ \\

The remarkable fact, however, is that : \\

(2.26)~~~ $ \begin{array}{l}
                    \mbox{There are uncountably many pairwise} \\
                    \mbox{disjoint "walkable worlds". And between} \\
                    \mbox{any two disjoint "walkable worlds" there} \\
                    \mbox{are uncountably many other ones.}
             \end{array} $ \\

Also : \\

(2.27)~~~ $ \begin{array}{l}
                    \mbox{Non-disjoint "walkable worlds" are nested} \\
                    \mbox{into one another in uncountable chains.} \\
             \end{array} $ \\

As for the complex manner in which the "walkable worlds" are situated outside of one another,
or on the contrary, nested within one another, the self-similarity property of $^*\mathbb{R}$
in subsection 2.3. can offer further insights. \\

\medskip
{\bf 2.6. The sizes and relative sizes of "walkable worlds"} \\

In order to give a further idea about the structure of $^*\mathbb{R}$, and in particular,
about the sizes and relative sizes of "walkable worlds", we mention the following well known
properties of numbers in $^*\mathbb{R}$. \\

Given any nonzero infinitesimal $\epsilon \in\, ^*\mathbb{R}$, that is, $\epsilon \in mon
( 0 ) \setminus \{~ 0 ~\}$, there exist nonzero infinitesimals $\delta, \eta \in mon ( 0 )
\setminus \{~ 0 ~\}$, such that $\epsilon / \delta$ is again an infinitesimal, while $\epsilon
/ \eta$ is infinitely large. \\
Similarly, given any nonzero infinitesimal $\epsilon \in mon ( 0 ) \setminus \{~ 0 ~\}$, there
exist nonzero infinitesimals $\delta\,', \eta\,' \in mon ( 0 ) \setminus \{~ 0 ~\}$, such that
$\delta\,' / \epsilon$ is again an infinitesimal, while $\eta\,' /\epsilon$ is infinitely
large. \\

Correspondingly, given any infinitely large $s \in\, ^*\mathbb{R}$, that is, $s \in\,
^*\mathbb{R} \setminus Gal ( 0 )$, there exist infinitely large $t,~ v \in\, ^*\mathbb{R}
\setminus Gal ( 0 )$, such that $s / t$ is infinitesimal, while $s / v$ is infinitely large. \\
Also, given any infinitely large $s \in\, ^*\mathbb{R} \setminus Gal ( 0 )$, there exist
infinitely large $t\,',~ v\,' \in\, ^*\mathbb{R} \setminus Gal ( 0 )$, such that $t\,' / s$ is
infinitesimal, while $v\,' / s$ is infinitely large. \\

Furthermore, given any two numbers $v,~ w \in\, ^*\mathbb{R},~ 0 < v \leq w$, then \\

(2.28)~~~ either ~~$ w / v $~~ is finite, or ~~$ w / v $~~ is infinitely large \\

which equivalently means that \\

(2.28$^*$)~~~ $ \begin{array}{l}
                    \mbox{either}~~ \exists~~ m \in \mathbb{N},~ m \geq 1 ~:~
                                                              w ~\leq~ n v \\ \\
                    ~~~~~\mbox{or}~~ \forall~~ m \in \mathbb{N},~ m \geq 1 ~:~ n v ~<~ w
                 \end{array} $ \\

As an effect of the above, we mention a few surprising facts about "walkable worlds". \\

First we recall that the Archimedean assumption, when considered within one dimension, has
ever since ancient times kept us - and still keeps much of Theoretical Physics - confined
within the real line $\mathbb{R}$ which is but a strict subset, see (2.3), (2.20), of the
"walkable world" given by the central galaxy, namely \\

(2.29)~~~ $ Gal ( 0 ) ~=~ WW_{1,\,0} $ \\

Compared to this ancient "only one walkable world" however, there are uncountably many
infinitesimal, or for that matter, infinitely large "walkable worlds". And on their turn,
infinitesimal "walkable worlds" can be infinitely large, or for that matter, infinitesimal
when compared with some other infinitesimal "walkable worlds". Similarly, infinitely large
"walkable worlds" can be infinitesimal, or on the contrary, infinitely large when compared
with some other infinitely large "walkable worlds". \\
In this regard, the central "walkable world" (2.29) has only one distinguishing feature,
namely, that it contains both 0 and 1. \\

Further, let us note that \\

(2.30)~~~ $ mon ( 0 ) $~~ is not a "walkable world" \\

although obviously $0 \in mon ( 0 )$. Indeed, let us assume that \\

$~~~~~~ mon ( 0 ) ~=~ WW_{\epsilon,\, \eta} $ \\

for a certain $\epsilon, \eta \in\, ^*\mathbb{R},~ \epsilon > 0$. But $0 \in mon ( 0 ) =
WW_{\epsilon,\, \eta}$\,, hence (2.19) gives \\

$~~~~~~ mon ( 0 ) ~=~ WW_{\epsilon,\, 0} $ \\

And then in view of (2.16$^*$), it follows that for every $\chi \in\, ^*\mathbb{R}$, there
exists $n \in \mathbb{N},~ n \geq 1$, such that \\

$~~~~~~ - n \epsilon ~\leq~ \chi ~\leq~  n \epsilon $ \\

which is obviously not true for $\chi = \sqrt \epsilon \in mon ( 0 )$. \\

In view of (2.30) it follows that $mon ( 0 )$ itself, that is, the set of infinitesimals, is
far more large and complex than being "one single walkable world". And this is to be expected
from its {\it self-similarity} property in (2.10), (2.11). \\ \\

{\bf 3. Non-Archimedean structures of higher dimensions} \\

As we have seen, the non-Archimedean nature of $^*\mathbb{R}$ can already bring with it a
surprisingly novel and rich structure. Needless to say, the structure of non-Archimedean
spaces of higher dimensions can only be yet more rich and surprising. \\

Such higher, in fact, infinite dimensional non-Archimedean spaces, in fact algebras of
generalized functions, have been successfully used in solving large classes of earlier
unsolved linear and nonlinear partial differential equations, see [10-33] and the references
cited there. \\

However, there are far simpler non-Archimedean algebras of {\it scalars} which can have a
relevant use in Theoretical Physics as replacements of the usual Archimedean fields
$\mathbb{R}$ or $\mathbb{C}$ of real, respectively, complex scalars, see [6-9] and the
references cited there. \\

What is to note with respect to such non-Archimedean algebras of scalars is that they are in
fact infinite dimensional. This is precisely why they are non-Archimedean. \\

On the other hand, their construction is perfectly similar with the traditional Cauchy-Bolzano
construction of the real numbers $\mathbb{R}$. And consequently, their use does not lead to
any additional difficulties, except for what still happens to be the surprising novelty and
richness of non-Archimedean structures. \\

The essence of the mentioned traditional construction of the real numbers $\mathbb{R}$ has
been formulated in the 1950s as the {\it reduced power} construction. And it is one of the
main instruments in Model Theory, which is a discipline within Mathematical Logic. \\
Fortunately, reduced powers can easily be constructed and used without any involvement of
Model Theory or Mathematical Logic. And they have actually been used in this way, for instance,
in constructing the completion of metric, normed, or uniform topological spaces. \\
Here for convenience, we recall the main features of the construction of reduced powers, see
[6-8] and the references cited there for further detail. \\

By the way, as noted earlier, reduced powers were extensively used in the construction of
large classes of algebras of generalized functions, as well as generalized scalars, both
employed in the solution of a large variety of linear and nonlinear PDEs, see [10-33]. \\

The mentioned large class of {\it algebras of scalars}, constructed as reduced powers, is
obtained as follows. \\

Let $\mathbb{K}$ be any algebra, among others the field $\mathbb{R}$ of real, or alternatively,
the field $\mathbb{C}$ of complex numbers, and let $\Lambda$ be any {\it infinite} set of
indices. Then we consider \\

(3.1) $~~~~~~ \mathbb{K}^\Lambda $ \\

which is the set of all functions $x : \Lambda \longrightarrow \mathbb{K}$, and which is
obviously an algebra with the point-wise operations on such functions. Furthermore, if
$\mathbb{K}$ is commutative or associative, the same will hold for the algebra
$\mathbb{K}^\Lambda$. \\

A possible problem with these algebras $\mathbb{K}^\Lambda$ is that they have {\it zero
divisors}, thus they are {\it not} integral domains, in the case of nontrivial $\mathbb{K}$,
that is, when $0,~ 1 \in \mathbb{K}$, and $1 \neq 0$. For instance, if we take $\mathbb{K} =
\mathbb{R}$ and $\Lambda = \mathbb{N}$, then \\

(3.2) $~~~~~~ \begin{array}{l}
       ( 1, 0, 0, 0,~.~.~.~ ),~ ( 0, 1, 0, 0,~.~.~.~ ) \in \mathbb{K}^\Lambda ~=~
                                       \mathbb{R}^\mathbb{N}, \\ \\
       ( 1, 0, 0, 0,~.~.~.~ ),~ ( 0, 1, 0, 0,~.~.~.~ ) ~\neq~
                                   0 \in \mathbb{K}^\Lambda ~=~ \mathbb{R}^\mathbb{N}
               \end{array} $ \\

and yet \\

(3.3) $~~~~~~ \begin{array}{l}
                       ( 1, 0, 0, 0,~.~.~.~ )~.~ ( 0, 1, 0, 0,~.~.~.~ ) ~=~
                             ( 0, 0, 0,~.~.~.~ ) ~=~ \\ \\
                        ~~~~=~ 0 \in \mathbb{K}^\Lambda ~=~ \mathbb{R}^\mathbb{N}
              \end{array} $ \\

This issue, however, can be dealt with in the following manner, to the extent that it may
prove not to be convenient. Let us take any {\it ideal} ${\cal I}$ in $\mathbb{K}^\Lambda$ and
construct the {\it quotient algebra} \\

(3.4) $~~~~~~ A ~=~ \mathbb{K}^\Lambda / {\cal I} $ \\

This quotient algebra construction has {\it four} useful features, namely

\begin{enumerate}

\item It allows for the construction of a large variety of algebras $A$ in (3.4),

\item The algebras $A$ in (3.4) are fields, if and only if the corresponding ideals ${\cal I}$
are {\it maximal} in $\mathbb{K}^\Lambda$,

\item The algebras $A$ in (3.4) are without zero divisors, thus are integral domains, if and
only if the corresponding ideals ${\cal I}$ are {\it prime} in $\mathbb{K}^\Lambda$,

\item If $\mathbb{K} = \mathbb{R}$, then there is a simple way to construct such ideals
${\cal I}$ in $\mathbb{K}^\Lambda$, due to their one-to-one correspondence with {\it filters}
on the respective infinite sets $\Lambda$.

\end{enumerate}

Here it should be mentioned that the above construction of quotient algebras in (3.4) has
among others a well known particular case, namely, the construction of the field
$^*\mathbb{R}$ of nonstandard reals, where one takes $\mathbb{K} = \mathbb{R}$, while
${\cal I}$ is a maximal ideal which corresponds to an ultrafilter on $\Lambda$. \\

Facts 2 and 3 above are a well known matter of undergraduate algebra. \\
Fact 4 is recalled here briefly for convenience. First we recall that each $x \in
\mathbb{R}^\Lambda$ is actually a function $x : \Lambda \longrightarrow \mathbb{R}$. Let us
now associate with each $x \in \mathbb{R}^\Lambda$ its {\it zero set} given by $Z ( x ) ~=~
\{~ \lambda \in \Lambda ~|~ x ( \lambda ) ~=~ 0 ~\}$, which therefore is a subset of
$\Lambda$. \\

Further, let us recall the concept of {\it filter} on the set $\Lambda$. A family ${\cal F}$
of subsets of $\Lambda$, that is, a subset ${\cal F} \subseteq {\cal P} ( \Lambda )$, is
called a {\it filter} on $\Lambda$, if and only if it satisfies the following three
conditions \\

(3.5) $~~~~~~ \begin{array}{l} 1.~~ \phi \notin {\cal F} \neq \phi \\ \\
                               2.~~ J,~ K \in {\cal F} ~~\Longrightarrow~~ J \bigcap K \in
                               {\cal F} \\ \\
                               3.~~ \Lambda \supseteq K \supseteq J \in {\cal F}
                                              ~~\Longrightarrow~~ K \in {\cal F}
                         \end{array} $ \\

Given now an ideal ${\cal I}$ in $\mathbb{R}^\Lambda$, let us associate with it the set of
zero sets of its elements, namely \\

(3.6) $~~~~~~ {\cal F}_{\cal I} ~=~ \{~ Z ( x ) ~~|~~ x \in {\cal I} ~\}
                                                   ~\subseteq~ {\cal P} ( \Lambda ) $ \\

Then \\

(3.7) $~~~~~~ {\cal F}_{\cal I} ~~\mbox{is a filter on}~ \Lambda $ \\

\medskip
Indeed, ${\cal I} \neq \phi$, thus ${\cal F}_{\cal I} \neq \phi$. Further, assume that
$Z ( x ) = \phi$, for a certain $x \in {\cal I}$. Then $x ( \lambda ) \neq 0$, for $\lambda
\in \Lambda$. Therefore we can define $y : \Lambda ~\longrightarrow~ \mathbb{R}$, by
$y ( \lambda ) = 1 / x ( \lambda )$, with $\lambda \in \Lambda$. Then however $y . x = 1 \in
\mathbb{R}^\Lambda$, hence ${\cal I}$ cannot be an ideal in $\mathbb{R}^\Lambda$, which
contradicts the hypothesis. In this way condition 1 in (3.5) holds for ${\cal F}_{\cal I}$. \\
Let now $x,~ y \in {\cal I}$, then clearly $x^2 + y^2 \in {\cal I}$, and $Z ( x^2 + y^2 ) =
Z ( x ) \bigcap Z ( y )$, thus condition 2 in  (3.5) is also satisfied by
${\cal F}_{\cal I}$. \\
Finally, let $x \in {\cal I}$ and $K \subseteq \Lambda$, such that $K \supseteq Z ( x )$. Let
$y$ be the characteristic function of $\Lambda \setminus K$. Then $x . y \in {\cal I}$, since
${\cal I}$ is an ideal. But now obviously $Z ( x . y ) = K$, which shows that
${\cal F}_{\cal I}$ satisfies as well condition 3 in (3.5). \\

There is also the {\it converse} construction. Namely, let ${\cal F}$ be any filter on
$\Lambda$, and let us associate with it the set of functions \\

(3.8) $~~~~~~ {\cal I}_{\cal F} ~=~ \{~ x : \Lambda ~\longrightarrow~ \mathbb{R} ~~|~~ Z ( x )
                                         \in {\cal F} ~\} ~\subseteq~ \mathbb{R}^\Lambda $ \\

Then \\

(3.9) $~~~~~~ {\cal I}_{\cal F} ~~\mbox{is an ideal in}~ \mathbb{R}^\Lambda $ \\

Indeed, for $x, y \in \mathbb{R}^\Lambda$, we have $Z ( x + y ) \supseteq Z ( x ) \bigcap
Z ( y )$, thus $x,~ y \in {\cal I}_{\cal F}$ implies that $x + y \in {\cal I}_{\cal F}$. Also
$Z ( x . y ) \supseteq Z ( x )$, therefore $x \in {\cal I}_{\cal F},~ y \in
\mathbb{R}^\Lambda$ implies that $x . y \in {\cal I}_{\cal F}$. Further we note that
$Z ( c x ) = Z ( x )$, for $c \in \mathbb{R},~ c \neq 0$. Finally, it is clear that
${\cal I}_{\cal F} \neq \mathbb{R}^\Lambda$, since $x \in {\cal I}_{\cal F} ~\Longrightarrow~
Z ( x ) \neq \phi$, as ${\cal F}$ satisfies condition 1 in (3.5).Therefore (3.9) does indeed
hold. \\

Let now ${\cal I},~ {\cal J}$ be two ideals in $\mathbb{R}^\Lambda$, while
${\cal F},~ {\cal G}$ are two filters on $\Lambda$. Then it is easy to see that \\

(3.10) $~~~~~~ \begin{array}{l}
                       {\cal I} ~\subseteq~ {\cal J} ~~~\Longrightarrow~~ {\cal F}_{\cal I}
                               ~\subseteq~ {\cal F}_{\cal J} \\ \\
                       {\cal F} ~\subseteq~ {\cal G} ~~~\Longrightarrow~~ {\cal I}_{\cal F}
                       ~\subseteq~ {\cal I}_{\cal G}
                \end{array} $ \\

We can also note that, given an ideal ${\cal I}$ in $\mathbb{R}^\Lambda$ and a filter
${\cal F}$ on $\Lambda$, we have by iterating the above constructions in (3.6) and (3.8) \\

(3.11) \quad $ \begin{array}{l}
                 {\cal I} ~~~\longrightarrow~~~ {\cal F}_{\cal I} ~~~\longrightarrow~~~
                             {\cal I}_{{\cal F}_{\cal I}} ~=~ {\cal I} \\ \\
                 {\cal F} ~~~\longrightarrow~~~ {\cal I}_{\cal F} ~~~\longrightarrow~~~
                             {\cal F}_{{\cal I}_{\cal F}} ~=~ {\cal F}
                        \end{array} $ \\

Indeed, in view of (3.6), (3.8), we have for $s \in \mathbb{R}^\Lambda$ the equivalent
conditions \\

$~~~~~~ x \in {\cal I} ~~~\Longleftrightarrow~~ Z ( x ) \in {\cal F}_{\cal I}
                  ~~~\Longleftrightarrow~~~ x \in {\cal I}_{{\cal F}_{\cal I}} $ \\

Further, for $J \subseteq \Lambda$, we have the equivalent conditions \\

$~~~~~~ J \in {\cal F}_{{\cal I}_{\cal F}} ~~~\Longleftrightarrow~~~ J ~=~ Z ( s ),
                                           ~~\mbox{for some}~ s \in {\cal I}_{\cal F} $ \\

But for $x \in \mathbb{R}^\Lambda$, we also have the equivalent conditions \\

$~~~~~~ x \in {\cal I}_{\cal F} ~~~\Longleftrightarrow~~~ Z ( x ) \in {\cal F} $ \\

and the proof of (3.11) is completed. \\

In view of (3.11), it follows that every ideal in $\mathbb{R}^\Lambda$ is of the form
${\cal I}_{\cal F}$, where ${\cal F}$ is a certain filter on $\Lambda$. Also, every filter on
$\Lambda$ is of the form ${\cal F}_{\cal I}$, where ${\cal I}$ is a certain ideal in
$\mathbb{R}^\Lambda$. \\ \\

{\bf 4. What about the Quanta and the Velocity of Light ?} \\

The {\it quantitative} aspect of the issue of the quanta and of the velocity of light has so
far made perfect sense within the Archimedean structure of the real line $\mathbb{R}$, that is,
within its "only one walkable world" view, being supported by countless physical
experiments. \\
Certainly, within such a view a {\it smallest allowed strictly positive} quantity, say, of
energy, and in that case called "energy quantum", can have a clear meaning. Indeed, outside of
that "only one walkable world" there is supposed to be nothing at all. Therefore, that quantum,
or smallest allowed strictly positive quantity - if it really exists - must by {\it necessity}
be situated within that very same "only one walkable world". And any number of physical
experiments support the existence of such quanta. \\
Similar is, of course, the situation with a {\it largest allowed finite} quantity, such as for
instance the velocity of light. \\

However, once that "only one walkable world" is replaced by a non-Archimedean structure, and
in fact, by {\it any} non-Archimedean one, a rather unprecedented richness of structure comes
into play due to the presence of {\it infinitesimals} and {\it infinitely large} elements, and
consequently, of the corresponding structure of "walkable worlds". \\
Consequently, the very meaning of smallest allowed strictly positive quantity, or
alternatively, of largest allowed finite quantity becomes {\it dependent} on the specific
"walkable world" within which it is considered. And as seen above even in the simple one
dimensional case of $^*\mathbb{R}$, such "walkable worlds" are uncountably many, either next
to one another, or nested within one another, all of them in a complex self-similar pattern.
In this way, it is no longer so clear within which particular "walkable worlds" should the
traditional concepts of quanta and velocity of light be considered. Thus such traditional
concepts may possibly need a fundamental reconsideration. \\

In this regard, here we conclude with a summary for an appropriate consideration of those of
the earlier presented facts in the simplest relevant non-Archimedean case, namely, of the one
dimensional $^*\mathbb{R}$, facts which may be relevant for the mentioned possible
reconsideration of quanta and/or velocity of light : \\

1) The self-similarity in (2.10), (2.11), namely

\begin{math}
\setlength{\unitlength}{0.2cm}
\thicklines
\begin{picture}(65,15)

\put(0,10){$[\, \bigcup_{r \in \mathbb{R},\, \lambda \in \Lambda}\,
                          ( r - s_\lambda + mon ( 0 ) ) \,] \cup
                   [\, \bigcup_{r \in \mathbb{R},\, \lambda \in \Lambda}
                          ( r + s_\lambda + mon ( 0 ) ) \,] \ni s ~\longmapsto$}
\put(2,3){$\longmapsto~ 1 / s \in (~ mon ( 0 ) \setminus \{~ 0 ~\} ~)$}

\end{picture}
\end{math}

and conversely

\begin{math}
\setlength{\unitlength}{0.2cm}
\thicklines
\begin{picture}(65,15)

\put(0,10){$(~ mon ( 0 ) \setminus \{~ 0 ~\} ~) \ni s \longmapsto$}
\put(2,3){$1 / s \in [\, \bigcup_{r \in \mathbb{R},\, \lambda \in \Lambda}\,
                          ( r - s_\lambda + mon ( 0 ) ) \,] \cup
                   [\, \bigcup_{r \in \mathbb{R},\, \lambda \in \Lambda}
                          ( r + s_\lambda + mon ( 0 ) ) \,]$}
\end{picture}
\end{math}

given by bijective order reversing mappings. \\

2) The order isomorphism between $Gal ( 0 )$ and any "walkable world" $WW_{u,\,s}$\,, with
$u,~ s \in\, ^*\mathbb{R},~ u > 0$, see (2.17). \\

3) The fact that $mon ( 0 )$ is {\it not} a "walkable world", but it has a rich structure made
up of them, see (2.30). \\

4) The fact that two "walkable worlds" are either disjoint, or one contains the other, see
(2.25). \\ \\

{\bf \large Appendix 1} \\

{\bf Basic Algebraic Structures} \\

For convenience, we recall here a few basic concepts from Algebra and Ordered Structures. The
respective concepts are introduced step by step, culminating with the ones we are interested
in, namely, {\it fields} and {\it algebras}, and their {\it Archimedean}, respectively, {\it
non-Archimedean} instances. \\

A {\it group} is a structure $( G, \alpha )$, where $G$ is a nonvoid set and $\alpha : G
\times G \longrightarrow G$ is a binary operation on $G$ which is :

\begin{itemize}

\item associative : \\ \\
      $ \forall~~ x, y, z \in G :
                          \alpha ( \alpha ( x, y ), z ) = \alpha ( x, \alpha ( y, z ) ) $

\item has a neutral element $e \in G$ : \\ \\
      $ \forall~~ x \in G :
                          \alpha ( x, e ) = \alpha ( e, x ) = x $

\item each element $x \in G$ has an inverse $x\,' \in G$ : \\ \\
     $ \alpha ( x, x\,' ) = \alpha ( x\,', x ) = e $

\end{itemize}

The group $( G, \alpha )$ is {\it commutative}, if and only if : \\

$ \forall~~ x, y \in G : \alpha ( x, y ) = \alpha ( y, x ) $ \\

In such a case the binary operation $\alpha$ is simply denoted by "$+$" and called {\it
addition}, namely \\

$ \alpha ( x, y ) = x + y,~~~ x, y \in G $ \\

Further, the neutral element is denoted by $0$, namely, $e = 0$, while for every $x \in G$,
its inverse $x\,'$ is denoted by $- x$. \\

It will be useful to note the following. Given any group element $x \in G$ and any integer
number $n \geq 1$, we can define the group element $n x \in G$, by \\

$ n x ~=~ \begin{array}{|l}
               ~~x ~~\mbox{if}~ n = 1 \\ \\
               ~~x + x + x + \ldots + x ~~\mbox{if}~ n \geq 2
          \end{array} $ \\

where the respective sum has $n$ terms. The meaning of this operation is easy to follow.
Namely, $n x$ can be seen as $n$ steps of length $x$ each, in the direction $x$. This
interpretation will be particularly useful in understanding the condition defining the
Archimedean property, and thus, of the non-Archimedean property as well. \\

We recall that the usual addition gives a commutative group structure on the integer numbers
$\mathbb{Z}$, rational numbers $\mathbb{Q}$, real numbers $\mathbb{R}$, complex numbers
$\mathbb{C}$, as well as on the set $\mathbb{M}^{\,m,\,n}$ of $m \times n$ matrices, for every
$m, n \geq 1$. \\

Now, a {\it ring} is a commutative group $( S, + )$ on which a second binary operation $\beta
: S \times S \longrightarrow S$, called multiplication, is defined with the properties :

\begin{itemize}

\item $\beta$ is associative

\item $\beta$ is distributive with respect to addition : \\ \\
      $\forall~~ x, y , z \in S : \\
         ~~~~~\beta ( x, y + z ) = \beta ( x, y ) + \beta ( x, z ),~~~
                                 \beta ( x + y, z ) = \beta ( x, z ) + \beta ( y, z ) $

\end{itemize}

Usually, this second binary operation $\beta$ is called {\it multiplication}, and it is
denoted by ".", namely \\

$ \beta ( x, y ) = x\,.\,y,~~~ x, y \in S $ \\

and often, it is denoted even simpler as merely $x y = x\,.\,y$, with $x, y \in S$. \\

The ring $( S, + , . )$ is called {\it unital}, if and only if there is an element $u \in S$,
such that \\

$ \forall~~ x \in S : u\,.\,x = x\,.\,u = x $ \\

Usually, the element $u \in S$ is denoted by $1$, namely \\

$ u = 1 $ \\

The ring $( S, + , . )$ is called {\it commutative}, if and only if \\

$ \forall~~ x, y \in S : x\,.\,y = y\,.\,x $ \\

We recall that with the usual addition and multiplication, the integer numbers $\mathbb{Z}$,
rational numbers $\mathbb{Q}$, real numbers $\mathbb{R}$ and complex complex numbers
$\mathbb{C}$ are commutative unital rings, while the set $\mathbb{M}^{\,m,\,n}$ of $m \times n$
matrices, with $m, n \geq 2$, are {\it noncommutative} unital rings. \\

An important concept in rings is that of {\it zero divisor}. Namely, two elements $x, y \in S$
are called zero divisors, if and only if

\begin{itemize}

\item $ ~~~~~~ x \neq 0,~~~ y \neq 0 $

\item $ ~~~~~~ x\,.\,y = 0 $

\end{itemize}

Clearly, $\mathbb{Q},~ \mathbb{R}$ and $\mathbb{C}$ are rings {\it without} zero divisors,
while the set $\mathbb{M}^{\,m,\,n}$ of $m \times n$ matrices, with $m, n \geq 2$, has zero
divisors, the latter fact being illustrated already in the case of $\mathbb{M}^{\,2,\,2}$ by
such a simple example as \\

$ \left ( \begin{array}{l}
                 1~~~~~~0 \\
                 0~~~~~~0
           \end{array} \right )
  \left ( \begin{array}{l}
                 0~~~~~~0 \\
                 0~~~~~~1
           \end{array} \right ) ~=~
  \left ( \begin{array}{l}
                 0~~~~~~0 \\
                 0~~~~~~0
           \end{array} \right )$ \\

Commutative rings without zero divisors are called {\it integral domains}. \\

The consequence of the above is that in rings with zero divisors one {\it cannot always}
simplify factors in a product. Namely, for $x, y \in S$, the relation \\

$ x\,.\,y = 0 $ \\

need {\it not always} imply that \\

$ x = 0~~~\mbox{or}~~ y = 0 $ \\

just as happens with the above product of two matrices. This further means that, given $x, y,
z \in S$, the relation \\

$ x\,.\,y = x\,.\,z $ \\

or for that matter, the relation \\

$ y\,.\,x = z\,.\,x $ \\

need {\it not always} imply that \\

$ y = z $ \\

eve if $x \neq 0$. \\

As an effect, in rings with zero divisors {\it not} every nonzero element has an inverse.
Indeed, assuming the contrary, let $x\,.\,y = 0$, with $x, y \in S,~ x \neq 0$. Then there
exists an inverse $x\,' \in S$ for $x$, which means that $x\,.\,x\,' = x\,.\,x\,' = 1$. Hence
$x\,'\,.\,( x\,.\,y) = x\,'\,.\,0$, or due to the associativity of the product, we have
$( x\,'\,.\,x )\,.\,y = 0$, which means $y = 1\,.\,y = ( x\,'\,.\,x )\,.\,y = 0$. Thus we
obtained that $x\,.\,y = 0$ and $x \neq 0$ imply $y = 0$, which gives the contradiction that
$S$ cannot have zero divisors. \\

An algebraic structure of great importance is that of {\it fields}. A ring $( F, +, . )$ is a
field, if and only if every nonzero element $x \in F$ has an inverse $x\,' \in F$, namely \\

$ x\,.\,x\,' = x\,'\,.\,x = 1 $ \\

It follows that a field {\it cannot} have zero divisors. \\

In this regard, $\mathbb{Q},~ \mathbb{R}$ and $\mathbb{C}$ are fields, while $\mathbb{Z}$ and
$\mathbb{M}^{\,m,\,n}$, with $m, n \geq 2$, are {\it not} fields. \\
The ring $\mathbb{Z}$ is, as we have seen, an integral domain. But it is not a field, since
none of its nonzero elements, except for $1$ and $-1 $, has an inverse. \\
On the other hand, as we have seen, the rings $\mathbb{M}^{\,m,\,n}$, with $m, n \geq 2$, have
zero divisors, thus they cannot be fields. \\

Lastly, a ring $( A, + , . )$ is called an {\it algebra} over a given field $\mathbb{K}$, if
and only if there exists a third binary operation $\gamma : \mathbb{K} \times A
\longrightarrow A$, called {\it multiplication with a scalar}, namely, for each scalar $a \in
\mathbb{K}$, and each algebra element $x \in A$, we have $\gamma ( a, x ) \in A$. \\
Usually, this binary operation $\gamma$ is also written as a multiplication ".", even if that
may cause confusion. However, one should remember that in an algebra there are {\it two} multiplications,
namely, one between two algebra elements $x, y \in A$, and which gives the algebra element
$x\,.\,y \in A$, and another multiplication between a scalar $a \in \mathbb{K}$ and an algebra
element $x \in A$, giving the algebra element $a\,.\,x \in A$. \\
The properties of this second binary operation, namely, of multiplication with scalars, are as
follows. For $a, b \in \mathbb{K},~ x, y \in A$, we have : \\

\begin{itemize}

\item ~~~~~~ $ a\,.\,( x + y ) = ( a\,.\,x ) + ( a\,.\,y ) $

\item ~~~~~~ $ ( a + b )\,.\,x = ( a\,.\,x ) + ( b\,.\,x ) $

\item ~~~~~~ $ ( a\,.\,b )\,.\,x = a\,.\,( b\,.\,x ) $

\item ~~~~~~ $ 1\,.\,x = x $

\end{itemize}

Given a field $\mathbb{K}$, for instance, $\mathbb{K} = \mathbb{R}$, or $\mathbb{K} =
\mathbb{C}$, a typical and important algebra over $\mathbb{K}$ is the set
$\mathbb{M}\,^{m,\,n}_\mathbb{K}$ of $m \times n$ matrices with elements which are scalars in
$\mathbb{K}$, where $m, n \geq 2$. \\
Here the difference between the two multiplications is obvious. The first multiplication is
that between two matrices in $A, B \in \mathbb{M}\,^{m,\,n}_\mathbb{K}$. The second
multiplication is that between a scalar $a \in \mathbb{K}$ and a matrix
$A \in \mathbb{M}\,^{m,\,n}_\mathbb{K}$. \\ \\

{\bf The Archimedean Property} \\

The Archimedean property, as much as the property of being non-Archimedean, is essentially
related to certain {\it algebraic} + {\it order} structures. The simplest way to deal with the
issue is to consider {\it ordered groups}. And in fact, we can restrict ourself to commutative
groups. \\

Commutative groups were defined above, therefore, here we briefly recall the definition of
{\it partial orders}. \\

A partial order $\leq$ on a nonvoid set $X$ is a binary relation $x \leq y$ between elements
$x, y \in X$, which has the following three properties :

\begin{itemize}

\item $\leq$ is reflexive : \\ \\
      $ \forall~~ x \in X : x \leq x $

\item $\leq$ is antisymmetric : \\ \\
      $ \forall~~ x, y \in X : x \leq y,~ y \leq x ~\Longrightarrow~ x = y $

\item $\leq$ is transitive : \\ \\
      $\forall~~ x, y, z \in X : x \leq y,~ y \leq z ~\Longrightarrow~ x \leq z $

\end{itemize}

In case we have the additional property \\

$ \forall~~ x, y \in X : ~~\mbox{either}~ x \leq y, ~~\mbox{or}~ y \leq x $ \\

then $\leq$ is called a {\it linear} order on $X$. \\

Given now a commutative group $( G, + )$, a partial order $\leq$ on $G$ is called {\it
compatible} with the group structure, if and only if : \\

$ \forall~~ x, y, z \in G : x \leq y ~\Longrightarrow~ x + z \leq y + z $ \\

A {\it partially ordered commutative group} is by definition a commutative group $( G, + )$
together with a compatible partial order $\leq$ on $G$. In such a case, for simplicity, we
shall use the notation \\

$ ( G, +, \leq ) $ \\

In particular, we have a {\it linearly ordered commutative group} when the compatible partial
order $\leq$ is linear. \\

It is easy to see that in the general case of partially ordered commutative group $( G, +,
\leq )$, the above condition of compatibility between the partial order $\leq$ and the group
structure can be simplified as follows : \\

$ x, y \geq 0 ~\Longrightarrow~ x + y \geq 0 $ \\

where $0 \in G$ is the neutral element in $G$. \\

We recall that $\mathbb{Z},~ \mathbb{Q}$ and $\mathbb{R}$ are commutative groups. It is now
easy to see that with the usual order relation $\leq$, each of them is a linearly ordered
commutative group. \\

Examples of partially ordered commutative groups which are {\it not} linearly ordered are easy
to come by. Indeed, let us consider the $n$-dimensional Euclidean space $\mathbb{R}^n$, with
$n \geq 2$. With the usual addition of its vectors, this space is obviously a commutative
group. We can now define on it the partial order relation $\leq$ as follows. Given two vectors
$x = ( x_1, x_2, x_3, \ldots , x_n ),~ y = ( y_1, y_2, y_3, \ldots , y_n ) \in \mathbb{R}^n$,
then we define $x \leq y$ coordinate-wise, namely \\

$ x \leq y ~\Longleftrightarrow~
        x_1 \leq y_1,~ x_2 \leq y_2,~ x_3 \leq y_3, \ldots , x_n \leq y_n $ \\

Then it is easy to see that this partial order is compatible with the commutative group on
$\mathbb{R}^n$, but it is {\it not} a linear order, when $n \geq 2$. Indeed, this can be seen
even in the simplest case of $n = 2$, if we take $x = ( 1, 0 )$ and $y = ( 0, 1)$, since then
we do not have either $x \leq y$, or $y \leq x$. \\

In particular, $\mathbb{C}$, as well as and $\mathbb{M}\,^{m,\,n}_\mathbb{R},~
\mathbb{M}\,^{m,\,n}_\mathbb{C}$, with $m \geq 2$ or $n \geq 2$, are partially and {\it not}
linearly ordered commutative groups. Indeed, when it comes to their group structure, each of
them can be seen as an Euclidean space. Namely $\mathbb{C}$ is isomorphic with $\mathbb{R}^2$,
$\mathbb{M}\,^{m,\,n}_\mathbb{R}$ is isomorphic with $\mathbb{R}^{m n}$, while
$\mathbb{M}\,^{m,\,n}_\mathbb{C}$ is isomorphic with $\mathbb{R}^{2 m n}$. \\

Finally, we can turn to the issue of being, or for that matter, of not being Archimedean. \\

A partially ordered commutative group $( G, + , \leq )$ is called {\it Archimedean}, if and
only if : \\

$ \exists~~ u \in G,~ u \geq 0 : \forall~~ x \in G,~ x \geq 0 :
                    \exists~~ n \in \mathbb{N},~ n \geq 1 : n u \geq x $ \\

Obviously, rings, algebras and fields each have, as far as their respective operations of
addition are concerned, a commutative group structure as part of their definition. And when a
partial order is defined to be compatible with the respective ring, algebra or field structure,
it will among other conditions be required to be compatible with the mentioned commutative
group structure of addition. \\

Consequently, the Archimedean condition on rings, algebras and fields can be defined
exclusively in terms of the partially ordered commutative group structure of their respective
operations of addition. \\ \\

{\bf \large Appendix 2} \\

{\bf Letter to the Editor of The Mathematical Intelligencer} \\

In the Summer 2006 issue of The Mathematical Intelligencer there are two reviews of Roger
Penrose's book "The Road to Reality" published in 2005. These two reviews recall quite clearly
the standard political ways of two party Anglo-Saxon democratic systems as understood and
trivialized by journals such as Newsweek, or rather, they may simply recall the "good cop -
bad cop" approach to criminals. One can wonder whether The Mathematical Intelligencer did that
by mistake, or on the contrary, finds it a matter of pride to try to implant such approaches
into science. And in the less than fortunate latter case, one can wonder why only two opposing
views were presented when commenting upon scientific facts like, say, 1 + 1 = 2 ? \\
Why not, indeed, three, or even more opposing views ? \\
After all, why should we not have some sort of circus in such rather arid realms like
mathematics ? \\

And now back to the mentioned two reviews. The first of them is shorter and quite sparse in
detail when it finds the book under review highly meritorious and readable. \\
The second review recalls an old Jewish story which goes as follows. Somewhere in Medieval
Europe, in a place with lingering anti-Semitism, an old Jew is brought to a court of law.
Before sentencing, the judge allows the poor man to make a last statement, and this is what he
has to say : Your Honour, I only whish I were judged by you as your predecessor did. Yes, when
it came to sentencing, he said "He is a Jew, but he is innocent". And now, I am afraid, Your
Honour may say "He is innocent, but he is a Jew" \ldots \\
Well, the second review does find quite a number of outstanding features in Penrose's latest
book, but all of that is totally and hopelessly drowned in a manifestly vicious overall
prejudiced attitude and judgement. \\
One can only wonder how a third, or perhaps, fourth and so on, review might have looked, had
The Mathematical Intelligencer decided to do one better than the trivial Newsweek approach,
and present us more than merely the rock bottom two sharply opposing views \ldots \\

And now, may I myself make a brief comment on Penrose's mentioned book, and start by noting
that, as it happens, I myself have had some conflicting arguments with him on certain strictly
mathematical issue, thus I cannot be counted as one of his unconditional admirers. \\
First, and above all, the subject of Penrose's latest book is by far the most fundamental and
consequential within the sciences of the last few centuries of our modern times. \\
Second, the way science is pursued for more than half a century by now, scholarship, and even
more so a wide ranging and deeply reaching one, is massively discouraged, in favour of
narrowly specialized research production. Penrose happens to be one of the very few scholars,
if not in fact the only one nowadays, with a truly impressive depth and breadth in the subject.
And in addition to having himself significant research contributions, he has clearly been one
of the rather rare breed of distinguished "thinking scientists", and not merely one of the
many many merely "working" ones. Consequently, even if his latest book were rather poor, which
clearly it is very far from being, one should nevertheless have an extraordinary appreciation
for his scholarship and willingness to make the considerable effort to bring that scholarship
into the public domain. In this regard it important to point out that it shows a very poor
understanding on the part of any reviewer or reader to see Penrose's latest book as a science
popularization one. Indeed, the kind of science it covers simply cannot be popularized too
widely. And it is due not only to the mathematics which a more general readership may lack,
but also, and in no less measure, to the extremely counterintuitive nature of much of modern
physics. \\

And then, Penrose's latest book is in fact like a Himalaya he built in the public domain, with
a grand and most fascinating view of that fundamental and all important field of science. And
that view may, hopefully, tempt many in the future young generations to try to climb that
wonder, each according to his or her ability. As for the rest who care to look at it, in this
case browse it or read parts of it, it may serve as one outstanding way to connect somewhat
better to things beyond, and no less important, than day to day concerns or events. \\
And the fact is that, when it comes to the quality of the book, the first reviewer's
appreciation is far more to the point, and it is only a pity that he did not elaborate on his
respective arguments in more abundant detail.

\end{document}